\documentclass[12pt]{iopart}
\usepackage{iopams}  
\usepackage{graphicx}
\usepackage[active]{srcltx}
\newcommand{\green}[2]{\ensuremath{\langle\!\langle #1;#2\rangle\!\rangle}}
\newcommand{\corr}[1]{\ensuremath{\langle #1 \rangle}}
\newcommand{\mean}[1]{\mathord{\left\langle #1 \right\rangle}}

\begin{document}

\title[Disordered Kondo-lattice model]{The ground state phase diagram of the diluted ferromagnetic Kondo-lattice model}

\author{M. Stier$^1$, S. Henning$^2$ and W. Nolting}

\address{Lehrstuhl Festk\"orpertheorie, Institut f\"ur Physik, Humboldt-Universit\"at zu Berlin, Newtonstrasse 15, 12489 Berlin, Germany}
\eads{\mailto{stier@physik.hu-berlin.de}$^1$, \mailto{henning@physik.hu-berlin.de}$^2$}
\begin{abstract}
We investigate the existence of several (anti-)ferromagnetic phases in the diluted ferromagnetic Kondo-lattice model, i.e. ferromagnetic coupling of local moment and electron spin. To do this we use a coherent potential approximation (CPA) with a dynamical alloy analogy. For the CPA we need effective potentials, which we get first from a mean-field approximation. To improve this treatment we use in the next step a more appropriate moment conserving decoupling approach and compare both methods. The different magnetic phases are modelled by defining two magnetic sublattices. As a result we present zero-temperature phase diagrams according to the important model parameters and different dilutions.
\end{abstract}

\maketitle

\section{Introduction}

In solid state physics magnetism is one of the most discussed phenomena. Especially for spintronic devices, which use in addition to the charge the spin of the electrons, it is necessary to understand the magnetism of highly correlated materials. One class of materials that could be used for spintronics are the diluted magnetic semiconductors (DMS) (\cite{ohno,akai,dmshightc,dietl2000}). These consist of semiconductors as a host material in which small amounts of magnetically active atoms are deposed. The ambition is that the resulting material has the properties of a semiconductor and is magnetic, too. But the mixture of materials can also destroy one or both qualities.\\
To describe DMS in theory several approaches are used. Much work is done based on ab initio calculations \cite{Katayama2003,sato2004,craco2003, larson1988,sato2010}. Those calculations can give material specific information and therefore show the chemical origin of differences between various semiconductors. In addition the exchange mechanism is usually described by a model Hamiltonian. It lies in the nature of a model to simplify the actually appearing physical processes and to concentrate on the most important mechanism occurring in a certain material. Thus diverse models are used to describe different DMS. These include, e.g., the double exchange\cite{sato2004}, superexchange\cite{sato2006} and/or the Kondo-lattice model (KLM)\cite{alvarez2002,alvarez2003,Berciu2001}.\\
Since the KLM is widely used for the description of DMS, we want to discuss its possible magnetic ground states. Doing this we want to compare the internal energy of different states. This, of course, implies a predetermination of those states. Besides the fully ordered ferromagnetic and the completely disordered paramagnetic state, regarding the local moments, there can be, e.g., several antiferromagnetic states, spin-canted ones, spirals, spin-density waves or even spin-glas phases. All of them were found to be ground states in specific models or parameter regions \cite{rev1997,prb1987,shen1999}. Since a complete description of those states is not possible in our model we will focus on five important types of possible ground states. This should yet show some general trends of the phase diagram of the diluted KLM. Zero temperature magnetic phase diagrams were already calculated for the concentrated KLM (\cite{pruschke,stier, Henning09,pd2009}), but, to our knowledge, not for the diluted one.\\
This work is organized as follows: First we show which magnetic phases are investigated in this paper and how they are described by a definition of magnetic sub-lattices. Then we use a equation of motion approach to get approximate solutions of the many-body problem of the Kondo-lattice model. This gives us effective potentials which we can include in a coherent potential approximation. After that we are able to calculate the internal energy of the single magnetic phases and present phase diagrams for different concentrations. The origin of those phases are discussed by means of quasi-particle density of states. Finally we summarize this work and give a short outlook to future work.

\section{Model and Theory}

\begin{figure}
	\begin{center}
		\includegraphics[width=12.0cm]{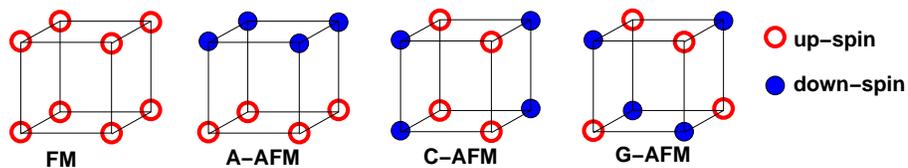}
		\caption{\label{fig:latticetypes}(Color online) Magnetic phases considered in this work.}
	\end{center}
\end{figure}

We want to investigate which type of magnetic order is present in diluted magnetic systems at zero temperature. In our work the magnetic component is described by a KLM with a positive coupling $J$, commonly known as ferromagnetic Kondo-lattice model (FKLM) (\cite{nolting2003,Mancini}), although the notion may be considered as somewht misleading. Since the carriers in this model study are electrons, this positive sign of $J$ is equivalent to a negative one in the hole picture\cite{kreissl2005} which is usually assumed for DMS. The FKLM is characterized by an interaction between itinerant electrons of spin $\boldsymbol\sigma_i$ and localized moments $\mathbf S_i$. Its Hamiltonian reads
\begin{eqnarray}
H & = & H_{s}+H_{sd}=t\sum_{\langle ij\rangle\sigma}c^+_{i\sigma}c_{j\sigma}
		-J\sum_{i}x_i\mathbf{S}_i\cdot\boldsymbol\sigma_i
\end{eqnarray}
Due to the interaction part $H_{sd}$ it is energetically favorable when the electron's and the local moment's spin on the same site $\mathbf R_i$ align parallel for positive $J$. The next-neighbor hopping $t$ of the itinerant electrons leads to an indirect coupling of the local moments on different sites. For our purposes it is better to represent the spin of the electrons by their creators/annihilators $c^{(+)}_{i\sigma}$ via $\boldsymbol\sigma_i = \frac{1}{2}\sum_{\sigma\sigma'}c^+_{i\sigma}\hat{\boldsymbol\sigma}_{\sigma\sigma'}c_{i\sigma'}$, where $\hat{\boldsymbol\sigma}$ is the vector of the three Pauli matrices. When we write down the scalar product $\mathbf{S}_i\cdot\boldsymbol\sigma_i$ explicitly we get
\begin{eqnarray}
H&=&t\sum_{\langle ij\rangle\sigma}c^+_{i\sigma}c_{j\sigma}
		-\frac{J}{2}\sum_{i\sigma}x_i\left(z_{\sigma}S^{z}_i c^+_{i\sigma}c_{i\sigma}+S^{\sigma}_ic^+_{i-\sigma}c_{i\sigma}\right)\ .
\label{eq:hamilton}
\end{eqnarray}
$S_i^z$ is the $z$-component of the $\mathbf S_i$ and $S^{\sigma}_i=S^x_i+z_{\sigma} i S^y_i$, with $z_{\sigma}=\delta_{\sigma\uparrow}-\delta_{\sigma\downarrow}$, are the ladder operators. To take into account disorder we use a site dependent parameter $x_i=0,1$ which turns the interaction on or off. The sum over these parameters is fixed by the given concentration $c=\frac{1}{N}\sum_i x_i$ of the magnetic atoms.\\
To include antiferromagnetic phases we have to divide the lattice into sub-lattices. In this work we only consider (anti-)ferromagnetic phases which can be divided into two magnetic sublattices (cf. figure \ref{fig:latticetypes}). Within each sublattice we can define a sub-lattice magnetization, where the magnetizations of the two sub-lattices for the antiferromagnetic phases are antiparallel to each other. We assume a ferromagnetically saturated sub-lattice magnetization at $T=0$.  Even though it is known that these Neel states are not the ground states of the model, due to spin fluctuations, they are energetically close to the real ground states, especially for large spins. To take the sub-lattice formulation into account we change the Hamiltonian (\ref{eq:hamilton}) to
\begin{equation}
 H = \sum_{\mean{i,j}\sigma\alpha\beta}t^{\alpha\beta}_{ij}c^+_{i\alpha\sigma}c_{j\beta\sigma} -\frac{J}{2}\sum_{i\sigma\alpha}x_{i\alpha}\left(z_{\sigma}S^{z}_{i\alpha} c^+_{i\sigma\alpha}c_{i\sigma\alpha}+S^{\sigma}_{i\alpha}c^+_{i-\sigma\alpha}c_{i\sigma\alpha}\right)\ .\label{eq:hamilton_trans}
\end{equation}
The lattice sites $\mathbf R_i,\mathbf R_j$ are now within a magnetic sublattice and are shifted by a primitive translation vector $\mathbf r_{\alpha}$ according to the specific sub-lattice.
Since we yet have next-neighbor hopping in the \emph{chemical} lattice the hopping within each sub-lattice is constricted. For example the next-neighbors of the magnetic and the chemical  lattice in the A-type phase are only the same in,e.g., $xy$-planes. Thus the intra-lattice hopping $t^{\alpha\alpha}_{ij}$ is only two-dimensional. The hopping in $z$-direction leads to a hybridization of the two sub-lattices. Fourier-transformation of the hoppings yields the dispersion of the free system and can be written for the A-type AFM as
\begin{eqnarray}
 \epsilon^{\alpha\alpha}(\mathbf k)&=& \epsilon^{-\alpha-\alpha}(\mathbf k)=\frac{W}{6}\left(\cos(ak_x)+\cos(ak_y)\right)\\
 \epsilon^{\alpha-\alpha}(\mathbf k)&=& \epsilon^{-\alpha\alpha}(\mathbf k)=\frac{W}{6}\cos(ak_z)
\end{eqnarray}
where $W=12t$ is the free bandwidth of the bulk system and $a$ the lattice constant. For the C-type the intra-lattice hopping is one-dimensional and in the G-type there is no hopping within the magnetic lattice.\\
The ferromagnetic phase needs no definition of sub-lattices, of course. But this could be also done artificially by choosing one of the lattice decompositions of the antiferromagnetic phases. The sub-lattice magnetizations are now parallel ($\mean{\mathbf S_{i\alpha}}=\mean{\mathbf S_{i\alpha'}}$) to each other. No matter which decomposition we choose, we get the same results as for the non-decomposed case. This is a strong confirmation of the validity of the sub-lattice method.

\subsection{Internal energy}

The internal energy $U$ of the FKLM at $T=0$ is given by the ground-state expectation value of the
Hamiltonian (\ref{eq:hamilton_trans}). A straightforward calculation using the spectral
theorem then shows, that $U$ can be calculated from the retarded local electronic-GF 
($G^{\alpha\beta}_{ij\sigma}(E)=\green{c_{i\sigma\alpha}}{c^+_{j\sigma\beta}}_E$):
\begin{equation}
U = \langle H \rangle = \sum_{i\sigma\alpha}\int^{\infty}_{-\infty}Ef_{-}(E)
							\left(-\frac{1}{\pi}\mathrm{Im}G^{\alpha\alpha}_{ii\sigma}(E)\right)\mathrm{d}E,
\label{eq:internal_E}
\end{equation}
where $f_{-}(E)$ denotes the Fermi function. Until now this is an exact formula but since (\ref{eq:hamilton_trans})
defines a complicated many-body problem approximations in calculating the electronic-GF have to be
accepted. Therefore we come now to the discussion of this point.

\subsection{Treatment of the many-body problem}

Our starting point is the equation of motion (EOM) of the electronic GF:
\begin{equation}
\sum_{l\alpha''}(E\delta^{\alpha\alpha''}_{il}-t^{\alpha\alpha''}_{il})G^{\alpha''\alpha'}_{lj\sigma}(E) 
	 =  \delta^{\alpha\alpha'}_{ij} - x_{i}\frac{J}{2}\left(z_{\sigma}I^{\alpha\alpha\alpha'}_{iij\sigma}(E)+F^{\alpha\alpha\alpha'}_{iij\sigma}(E) 
	 	\right),
\label{eq:EQM}
\end{equation}
with Ising-GF
\begin{equation}
 I^{\alpha\alpha\alpha'}_{ikj\sigma}(E)=\green{S^z_{i\alpha} c_{k\sigma\alpha}}{c^+_{j\sigma\alpha'}}
\end{equation}
and spin-flip-GF (SF-GF)
\begin{equation}
 F^{\alpha\alpha\alpha'}_{ikj\sigma}(E)=\green{S^{-\sigma}_{i\alpha}c_{k-\sigma\alpha}}{c^+_{j\sigma\alpha'}}\ .
\end{equation}
The main idea for construction of the phase-diagram is to assume certain configurations for the local moment system and then to compare their respective internal
energies. Under the assumption of saturation the Ising-GF can be decoupled directly: 
$z_{\sigma}\green{S^z_{i\alpha}c_{i\sigma\alpha}}{c^+_{j\sigma\alpha'}}\rightarrow z_{\sigma}SG^{\alpha\alpha'}_{ij\sigma}(E)$.
In \cite{Henning09} we have shown, that the influence of the SF-GF on the ground state phase-diagram is small in the case of a system without dilution, at least if we only consider phases which are ferromagnetically saturated in each sub-lattice. In a first attempt to solve (\ref{eq:EQM}) we thus neglect the SF-GF arriving at:
\begin{equation}
\sum_{l\alpha''}\left( (E+x_iz_{\alpha}z_{\sigma}\frac{J}{2}S)\delta^{\alpha\alpha''}_{il} 
					   - t^{\alpha\alpha''}_{il}\right)G^{\alpha''\alpha'}_{lj\sigma(\mathrm{MF})}(E)
					   =\delta^{\alpha\alpha'}_{ij}.
\label{eq:MF_ocpa}
\end{equation}
 We will call this solution the $T=0$ mean-field
(MF) solution.\\
To investigate the influences of spin-flip processes we will also perform a moment conserving decoupling approach (MCDA) as in Ref. \cite{Nolting97}. One advantage of this method is, that we are not restricted to ferromagnetic saturation within the sub-lattices. Thus we can also specify the paramagnetic phase by setting $\mean{S^z_i}=0$. We formally define a self-energy in (\ref{eq:EQM}) which reads as
\begin{eqnarray}
 \sum_{l\alpha''}(E\delta^{\alpha\alpha''}_{il}-t^{\alpha\alpha''}_{il})G^{\alpha''\alpha'}_{lj\sigma}(E) &=& \delta^{\alpha\alpha'}_{ij\sigma}+\green{[c_{i\sigma\alpha},H_{sd}]_-}{c^+_{j\sigma\alpha'}}\nonumber\\
	 &=&  \delta^{\alpha\alpha'}_{ij\sigma} + x_{i}\sum_{l\alpha''}M^{\alpha\alpha''}_{il\sigma}(E)G^{\alpha''\alpha'}_{lj\sigma}(E) .\label{eq:EQM2}
\end{eqnarray}
The self-energy will play the role of a random potential later on, but we will first focus on a concentrated lattice ($x_i=1 \forall i$) to derive the formula for this self-energy. We assume $M^{\alpha\alpha''}_{il\sigma}(E)$ to be local which will not give rise to any contradiction in our treatment \cite{Nolting97}. Thus we set $M^{\alpha\alpha''}_{il\sigma}(E) = M_{\sigma\alpha}(E)\delta^{\alpha\alpha''}_{il}$ and have the simplified EOM of the concentrated lattice
\begin{equation}
 \sum_{l\alpha''}\left((E-M_{\alpha\sigma}(E))\delta^{\alpha\alpha''}_{il}-t^{\alpha\alpha''}_{il}\right)G^{\alpha''\alpha'}_{lj\sigma}(E) 
	 =  \delta^{\alpha\alpha'}_{ij} \label{eq:EQM3}\ .
\end{equation}
When we compare (\ref{eq:EQM}) and (\ref{eq:EQM2}) we find a relation of the self-energy and the higher Green's functions
\begin{eqnarray}
 M_{\alpha\sigma}(E)G_{ij\sigma}^{\alpha\alpha'}(E) &=&- \frac{J}{2}\left(z_{\sigma}I^{\alpha\alpha\alpha'}_{iij\sigma}(E)+F^{\alpha\alpha\alpha'}_{iij\sigma}(E) 
	 	\right)\nonumber\\
&=&-\frac{J}{2}\left(z_{\sigma}\mean{S_{i\alpha}^z}G^{\alpha\alpha'}_{ij\sigma}(E)+z_{\sigma}\Gamma^{\alpha\alpha\alpha'}_{iij\sigma}(E)+F^{\alpha\alpha\alpha'}_{iij\sigma}(E) 
	 	\right) \label{eq:self_IF}
\end{eqnarray}
with the reduced Ising function $\Gamma^{\alpha\alpha\alpha'}_{iij\sigma}(E) = I^{\alpha\alpha\alpha'}_{iij\sigma}(E) -\mean{S_{i\alpha}^z}G^{\alpha\alpha'}_{ij\sigma}(E)=\green{\delta S^z_i c_{i\sigma\alpha}}{c^+_{j\alpha\sigma'}}$. Furthermore we read off from (\ref{eq:EQM2}) a formal correlation between self-energy and commutator
\begin{equation}
 \left[c_{i\sigma\alpha},H_{sd}\right] \longrightarrow M_{\alpha\sigma}(E) c_{i\sigma\alpha}\ ,\label{eq:comm_self}
\end{equation}
being rigorous, of course, only within the respective Green's functions. Now we write down the EOM for the higher GFs
\begin{eqnarray}
\fl \sum_{l\alpha''}\left( E\delta^{\alpha\alpha''}_{kl}-t^{\alpha	\alpha''}_{kl}\right)
						&F^{\alpha\alpha''\alpha'}_{ilj\sigma}(E) =  \nonumber\\
&	\green{[S^{-\sigma}_{i\alpha},H_{sd}]_{-}c_{k\alpha-\sigma}}{c^{+}_{j\alpha'\sigma}} + 
	\green{S^{-\sigma}_{i\alpha}[c_{k\alpha-\sigma},H_{sd}]_{-}}{c^{+}_{j\alpha'\sigma}}\ ,
\label{eq:SF_EQM}\\
\fl \sum_{l\alpha''}\left( E\delta^{\alpha\alpha''}_{kl}-t^{\alpha\alpha''}_{kl}\right)
						&\Gamma^{\alpha\alpha''\alpha'}_{ilj\sigma}(E) =  \nonumber\\
&	\green{[\delta S^{z}_{i\alpha},H_{sd}]_{-}c_{k\alpha\sigma}}{c^{+}_{j\alpha'\sigma}} + 
	\green{\delta S^{z}_{i\alpha}[c_{k\alpha\sigma},H_{sd}]_{-}}{c^{+}_{j\alpha'\sigma}}.
\label{eq:rI_EQM}
\end{eqnarray}
We will perform  different treatments of the non-local ($i\neq k$) and the local ($i=k$) terms in these equations. Let us start with the non-local ones. It has been shown in \cite{Nolting97} that the commutators containing a spin operator are connected to magnon energies. Those are typically several orders of magnitude smaller than the electron energy. Thus we neglect these commutators. For the other ones we assume that we can replace them by the formal expression (\ref{eq:comm_self}).\\
This procedure is however inappropriate for the local terms due to the intra-atomic interaction between electron and localized spin on the same lattice site. In this case we consider the commutators explicitly. Equations (\ref{eq:SF_EQM}) and (\ref{eq:rI_EQM}) now read
\begin{eqnarray}
\fl \sum_{l\alpha''}\Big( (E-M_{-\sigma\alpha}(E))\delta^{\alpha\alpha''}_{kl}&-t^{\alpha\alpha''}_{kl}\Big)
						F^{\alpha\alpha''\alpha'}_{ilj\sigma}(E) =  \nonumber\\
	&\delta^{\alpha\alpha'}_{ik}\left(-M_{-\sigma\alpha}(E)F^{\alpha\alpha\alpha'}_{iij\sigma}(E)+\green{[S^{-\sigma}_{i\alpha}c_{i-\sigma\alpha},H_{sd}]_{-}}{c^{+}_{j\sigma\alpha'}} \right),\\
\fl \sum_{l\alpha''}\Big( (E-M_{\sigma\alpha}(E))\delta^{\alpha\alpha''}_{kl}&-t^{\alpha\alpha''}_{kl}\Big)
						\Gamma^{\alpha\alpha''\alpha'}_{ilj\sigma}(E) =  \nonumber\\
	&\delta^{\alpha\alpha'}_{ik}\left(-M_{\sigma\alpha}(E)\Gamma^{\alpha\alpha\alpha'}_{iij\sigma}(E)+\green{[\delta S^{z}_{i\alpha}c_{i\sigma\alpha},H_{sd}]_{-}}{c^{+}_{j\sigma\alpha'}} \right);
\end{eqnarray}
where we avoided double counting of the self-energy for the local terms by subtracting it on the right-hand-side. Multiplying $\sum_{k\alpha}G^{\alpha'\alpha}_{ik\pm\sigma}(E)$ to the according equation leads with (\ref{eq:EQM3}) to
\begin{eqnarray}
\fl F^{\alpha \alpha \alpha' }_{iij\sigma}(E)& = 
	G^{\alpha\alpha}_{ii-\sigma}(E)\left(-M_{-\sigma\alpha}(E)F^{\alpha\alpha\alpha' }_{iij\sigma}(E)+\green{[S^{-\sigma}_{i\alpha}c_{i-\sigma\alpha},H_{sd}]_{-}}{c^{+}_{j\sigma\alpha' }} \right)\label{eq:Fcomm}\\
\fl\Gamma^{\alpha \alpha \alpha' }_{iij\sigma}(E)& = 
	G^{\alpha\alpha}_{ii\sigma}(E)\left(-M_{\sigma\alpha}(E)\Gamma^{\alpha\alpha\alpha'}_{iij\sigma}(E)+\green{[\delta S^{z}_{i\alpha}c_{i\sigma\alpha},H_{sd}]_{-}}{c^{+}_{j\sigma\alpha' }} \right)\label{eq:Gcomm}\ .
\end{eqnarray}
The remaining commutators create various Green's functions. Additionally to the known ones we get four higher GF
\begin{eqnarray*}
 F^{\alpha\alpha'(1)}_{iiij\sigma}(E) &= \green{ S_{i\alpha}^{-\sigma}S_{i\alpha}^z c_{i-\sigma\alpha}}{c_{j\sigma\alpha'}^+},\\
 F^{\alpha\alpha'(2)}_{iiij\sigma}(E) &= \green{\delta (S_{i\alpha}^{-\sigma}S_{i\alpha}^{\sigma})c_{i\sigma\alpha}}{c_{j\sigma\alpha'}^+},\\
 F^{\alpha\alpha'(3)}_{iiiij\sigma}(E) &= \green{S_{i\alpha}^{-\sigma}n_{i\sigma\alpha}c_{i-\sigma\alpha}}{c_{j\sigma\alpha'}^+},\\
 F^{\alpha\alpha'(4)}_{iiiij\sigma}(E) &= \green{S_{i\alpha}^z n_{i-\sigma\alpha}c_{i\sigma\alpha}}{c_{j\sigma\alpha'}^+}.
\end{eqnarray*}
In ferromagnetic saturation we can express two of them exactly 
\begin{eqnarray}
 F^{\alpha\alpha'(1)}_{iiij\sigma}(E) &= \left(S-\frac{1}{2} +\frac{1}{2} z_{\sigma}\right)F^{\alpha\alpha\alpha'}_{iij\sigma}(E),\\
 F^{\alpha\alpha'(2)}_{iiij\sigma}(E) &= 0\ .\label{F2limit}
\end{eqnarray}
This cannot be done for $ F^{\alpha\alpha'(3,4)}_{iiiij\sigma}(E)$ but we know that in the limiting case of an empty conduction band both functions vanish and for a completely filled band we find
\begin{eqnarray}
  F^{\alpha\alpha'(3)}_{iiiij\sigma}(E)&= F^{\alpha\alpha\alpha'}_{iij\sigma}(E)\\
 F^{\alpha\alpha'(4)}_{iiiij\sigma}(E) &= \Gamma^{\alpha\alpha\alpha'}_{iij\sigma}(E)+\mean{S^z_i} G^{\alpha\alpha'}_{ij\sigma}(E)\ .
\end{eqnarray}
Thus we make the ansatz that these functions can be expressed for all electron densities $n$  and magnetizations $\mean{S^z}$ by the linear combinations
\begin{eqnarray}
  F^{\alpha\alpha'(\nu)}_{iiiij\sigma}(E)& = \alpha^{(\nu)}_{\alpha\sigma} G^{\alpha\alpha'}_{ij\sigma}(E) + \beta^{(\nu)}_{\alpha\sigma} F^{\alpha\alpha\alpha'}_{iij\sigma}(E) &,\quad\nu=1,3\label{F3lin}\\
  F^{\alpha\alpha'(\mu)}_{iiiij\sigma}(E)& = \alpha^{(\mu)}_{\alpha\sigma} G^{\alpha\alpha'}_{ij\sigma}(E) + \beta^{(\mu)}_{\alpha\sigma}  \Gamma^{\alpha\alpha\alpha'}_{iij\sigma}(E)&,\quad\mu=2,4\label{F4lin}\ .
\end{eqnarray}
We fix the coefficients by the requirement that the zeroth spectral moments of the functions are conserved
\begin{eqnarray}
 \fl M^{(0)}(  F^{\alpha\alpha'(\nu)}_{iiiij\sigma}(E))&\stackrel{!}{=}\alpha^{(\nu)}_{\alpha\sigma}M^{(0)}(G^{\alpha\alpha'}_{ij\sigma}(E))+\beta^{(\nu)}_{\alpha\sigma}M^{(0)}(X^{\alpha\alpha\alpha'}_{iij\sigma}(E))&,\quad \nu=1,2,3,4\label{eqM0}
\end{eqnarray}
with $X=F,\Gamma$. To determine $\alpha^{(\nu)}_{\alpha\sigma},\beta^{(\nu)}_{\alpha\sigma}$ we need additional conditions. For the first two functions we get these from the next moments
\begin{eqnarray}
 \fl M^{(1)}(  F^{\alpha\alpha'(\nu)}_{iiiij\sigma}(E))&\stackrel{!}{=}\alpha^{(\nu)}_{\alpha\sigma}M^{(1)}(G^{\alpha\alpha'}_{ij\sigma}(E))+\beta^{(\nu)}_{\alpha\sigma}M^{(1)}(X^{\alpha\alpha\alpha'}_{iij\sigma}(E))&,\quad \nu=1,2\ .
\end{eqnarray}
 Furthermore we can calculate expection values via the spectral theorem from the different GF, which should also fullfill (\ref{F3lin}) and (\ref{F4lin}). Replacing the GF by the according expectation values yields
\begin{eqnarray}
 F^{3}:\ \mean{S_{i\alpha}^{-\sigma}c_{i\alpha\sigma}^+n_{i\alpha\sigma}c_{i\alpha-\sigma}} &=0= \alpha^{(3)}_{\alpha\sigma}\mean{n_{i\alpha\sigma}}+\beta^{(3)}_{\alpha\sigma}\mean{S^{-\sigma}_{i\alpha}c^+_{i\alpha\sigma}c_{i\alpha-\sigma}}\\
 F^{4}:\ \mean{S_{i\alpha}^{z}c_{i\alpha\sigma}^+n_{i\alpha-\sigma}c_{i\alpha\sigma}} &= \alpha^{(4)}_{\alpha\sigma}\mean{n_{i\alpha\sigma}}+\beta^{(4)}_{\alpha\sigma}\mean{\delta S^{z}_{i\alpha}n_{i\alpha\sigma}}\nonumber\\
&\stackrel{!}{=}\alpha^{(4)}_{\alpha-\sigma}\mean{n_{i\alpha-\sigma}}+\beta^{(4)}_{\alpha-\sigma}\mean{\delta S^{z}_{i\alpha}n_{i\alpha-\sigma}}\label{eqF4evcons}\ .
\end{eqnarray}
The first expectation value vanishes due to the combination of the electronic operators and the other is spin invariant. Equation (\ref{eqF4evcons}) seems to lead to no solution, due to underdetermination of the parameters. But (\ref{eqM0}) already yields $\alpha^{(4)}_{\alpha\sigma}=\mean{S^z_{i\alpha}n_{i\alpha-\sigma}}$, since the zero-moment of the spin-flip function $F_{iij\sigma}^{\alpha\alpha\alpha'}(E)$ disappears. Thus, as the only general choice for the other parameter, remains $\beta^{(4)}_{\alpha\sigma}=\mean{n_{i\alpha-\sigma}}$. The explicit terms of the other coefficients can be found in \ref{app:coeff}. With these conditions we can trace back all appearing Green's functions to a set of only three GF. Now we get a linear system of equations from (\ref{eq:Fcomm},\ref{eq:Gcomm}) as 
\begin{eqnarray}
\hat A_{\alpha\sigma} \left(\begin{array}{l} F^{\alpha\alpha\alpha'}_{iij\sigma}(E)\\ \Gamma^{\alpha\alpha\alpha'}_{iij\sigma}(E) \end{array}\right) = -\frac{J}{2} G^{\alpha\alpha'}_{ij\sigma}(E)\left(\begin{array}{l} A_{F\alpha\sigma}\\ A_{\Gamma\alpha\sigma} \end{array}\right)\ , \label{eq:linsoe}
\end{eqnarray}
with the coefficient matrix
\begin{equation*}
\fl \hat A_{\alpha\sigma} =\left(
\begin{array}{cc}
1+ G^{\alpha\alpha}_{ii-\sigma}(E)(M_{\alpha-\sigma}(E)+\frac{J}{2}C_{F\alpha\sigma})& \frac{J}{2}B_{F\alpha\sigma}G^{\alpha\alpha}_{ii-\sigma}(E)  \\ 
 \frac{J}{2}C_{\Gamma\alpha\sigma}G^{\alpha\alpha}_{ii\sigma}(E)& 1+ G^{\alpha\alpha}_{ii\sigma}(E)(M_{\alpha\sigma}(E)+\frac{J}{2}B_{\Gamma\alpha\sigma})\\ 
               \end{array}
\right)\ .
\end{equation*}
The coefficients ($A_{F\alpha\sigma}, B_{F\alpha\sigma},\dots$) are explained in \ref{app:coeff}. Within the matrix there are only local terms which are not site dependent when we assume translational invariance within the sub-lattice. Therefore we can write the solutions of (\ref{eq:linsoe}) as
\begin{eqnarray}
 F^{\alpha\alpha\alpha'}_{iij\sigma}(E) &= F^{\alpha}_{\sigma}(E) G^{\alpha\alpha'}_{ij\sigma}(E)\ ,\\
 \Gamma^{\alpha\alpha\alpha'}_{iij\sigma}(E) &= \Gamma^{\alpha}_{\sigma} (E) G^{\alpha\alpha'}_{ij\sigma}(E)\ .
\end{eqnarray}
The only dependence on the lattice indices is in $G^{\alpha\alpha'}_{ij\sigma}(E)$. Comparing this with (\ref{eq:self_IF}) yields the self-energy
\begin{equation}
 M_{\alpha\sigma}(E) = -\frac{J}{2}\left(z_{\sigma}\mean{S^z_{i\alpha}} +z_{\sigma}\Gamma^{\alpha}_{\sigma}(E)+F^{\alpha}_{\sigma}(E)\right)\ ,\label{eq:selfMCDA}
\end{equation}
which is indeed local. In an ordered system we would put the self-energy into formula (\ref{eq:EQM3}) and had a closed set of equations. For a disordered system we will change this procedure a bit as described in the next section.

\subsection{Inclusion of disorder - CPA}
The formulas derived in the previous chapter still depend on the concrete placement of the local moments. 
For a macroscopic large system the physical properties should be rather described by averaged quantities 
as long as the system is far from critical behavior near phase transitions. 
Therefore our goal is to find the configurationally averaged GF: $\corr{G^{\alpha\alpha'}_{ij\sigma}(E)}$. The problem is
solved if we are able to find the self-energy $\Sigma^{\mathrm{eff}}_{\alpha\sigma}(E)$ of an effective 
medium with the according symmetry of the considered type of magnetism. Its general formula is
\begin{eqnarray}
G^{\mathrm{eff},\alpha\alpha'}_{\mathbf{q}\sigma}(E)=\corr{G^{\alpha\alpha'}_{\mathbf{q}\sigma}(E)}
			=\left( (E-\Sigma^{\mathrm{eff}}_{\alpha\sigma})\delta_{\alpha\alpha'}
			-\epsilon^{\alpha\alpha'}({\mathbf{q}})\right)^{-1}_{\alpha\alpha'}.
\label{eq:effectM_self}
\end{eqnarray}
Let us inspect the MF-GF (\ref{eq:MF_ocpa}) first. The disorder problem is defined by a random change
of the local potential $\eta_{\sigma}$ for magnetic (M) and non-magnetic (NM) sites:
$\eta^{\mathrm{M}}_{\alpha\sigma} = -z_{\alpha}z_{\sigma}\frac{J}{2}S;\;\eta^{\mathrm{NM}}_{\alpha\sigma}=0$ with concentration $c$ and $1-c$ 
respectively. This is analogous to a binary alloy (for each direction of electron spin) with local disorder. 
We have applied the coherent potential approximation (CPA)\cite{Elliott74} for calculation of the effective medium self-energy. 
Since we are only interested in the diagonal parts of the GF $G^{\mathrm{eff},\alpha\alpha}_{\mathbf{q}\sigma}(E)$ and because of the locality of CPA self-energies, we can use a standard derivation of the CPA. One obtains the following (self-consistent) equation of a \textit{local} self-energy
approximation for the effective medium:
\begin{eqnarray}
0 = c	 \frac{\eta^{\mathrm{M}}_{\alpha\sigma}-\Sigma^{\mathrm{eff}}_{\alpha\sigma}}
		      {1-(\eta^{\mathrm{M}}_{\alpha\sigma}-\Sigma^{\mathrm{eff}}_{\alpha\sigma})G^{\mathrm{eff},\alpha\alpha}_{ii\sigma}}
  + (1-c)\frac{\eta^{\mathrm{NM}}_{\alpha\sigma}-\Sigma^{\mathrm{eff}}_{\alpha\sigma}}
		      {1-(\eta^{\mathrm{NM}}_{\alpha\sigma}-\Sigma^{\mathrm{eff}}_{\alpha\sigma})G^{\mathrm{eff},\alpha\alpha}_{ii\sigma}}.
\label{eq:cpa_selfconsistent}
\end{eqnarray}
Due to (\ref{eq:effectM_self}) the effective self-energies $\Sigma^{\mathrm{eff}}_{\alpha\sigma}(E)$ of the two sub-lattices are not independent of each other.\\
To improve the CPA result we now choose the self-energy of the MCDA $M_{\alpha\sigma}(E)=\eta_{\alpha\sigma}(E)$ in (\ref{eq:selfMCDA}) as an effective random potential. This potential is energy dependent, that is why we call this a dynamical alloy analogy (DAA, \cite{takahashi, tang}). But we have to keep in mind that self-energy was derived for a concentrated system. In a diluted system the magnetic subsystem hybridizes with the whole system.\\
Let us assume that we have found a solution for the self-energy $\Sigma^{\mathrm{eff}}_{\sigma}(E)$ of the diluted system. Now the Green's functions of the subsystems can be derived via the projection \cite{Velicky68}
\begin{equation}
 G^{\nu,\alpha\alpha}_{ii\sigma}(E) = \frac{G^{\mathrm{eff},\alpha\alpha}_{ii\sigma}(E) }{1- (\eta^{\nu}_{\alpha\sigma}(E) -\Sigma^{\mathrm{eff}}_{\alpha\sigma}(E))G^{\mathrm{eff},\alpha\alpha}_{ii\sigma}(E)}\label{projcpa},\qquad  \nu = \mathrm{M,NM}\ .\label{eq:cpaproj}
\end{equation}
We choose as an approximation that the GF of the magnetic subsystem $G^{\mathrm{M}\alpha\alpha'}_{ij\sigma}(E)$ can be treated as an effective concentrated system which has the same EOM as in a real concentrated system. Thus the formal structure remains the same and we just have to replace $G^{\pm\alpha\pm\alpha}_{ii\sigma}(E)$ in (\ref{eq:linsoe}) by $G^{\mathrm{M}\pm\sigma\pm\sigma}_{ii}(E)$. The self-energy and the magnetic sub-system's higher Green's functions ($\Gamma^{\mathrm{M}\alpha\alpha\alpha'}_{iij\sigma}(E), F^{\mathrm{M}\alpha\alpha\alpha'}_{iij\sigma}(E)$) are now functions of the projected Green's functions $G^{\mathrm{M}\pm\alpha\pm\alpha}_{ii\sigma}(E)$. To take into account the hybridization effects with the complete system we do not put the self-energy $M_{\alpha\sigma}(E)$ directly into formula (\ref{eq:EQM2}) to get $G^{\mathrm{M}\alpha\alpha}_{ij\sigma}(E)$ as we would do it in a concentrated lattice. Instead, we use it to get the effective self-energy $\Sigma^{\mathrm{eff}}_{\alpha\sigma}(E)$ defined in Eq. (\ref{eq:cpa_selfconsistent}). With $\Sigma^{\mathrm{eff}}_{\alpha\sigma}(E)$ we calculate $G^{\mathrm{eff}\alpha\alpha}_{ii\sigma}(E)$ and the projected Green's function (\ref{eq:cpaproj}) and the whole set of equations is closed.
\begin{figure}
 \begin{center}
  \includegraphics[width=.5\linewidth]{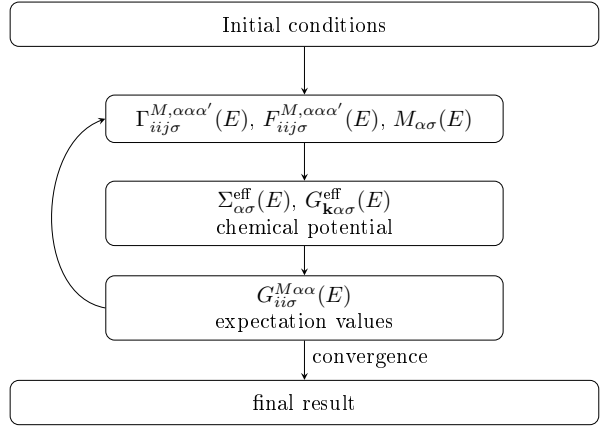}
 \end{center}
\caption{\label{fig:selfcons} The self-consistency cycle of the combined MCDA/CPA treatment. The magnetic sub-system (index "M") and the whole diluted system (index "eff") are not calculated separately, but in a combined loop.}
\end{figure}
This self-consistency cycle is depicted in figure \ref{fig:selfcons}.

\section{Results and Discussion}

\subsection{Mean-field results without dilution}

\begin{figure}
    \begin{center}
	\includegraphics[width=.6\linewidth]{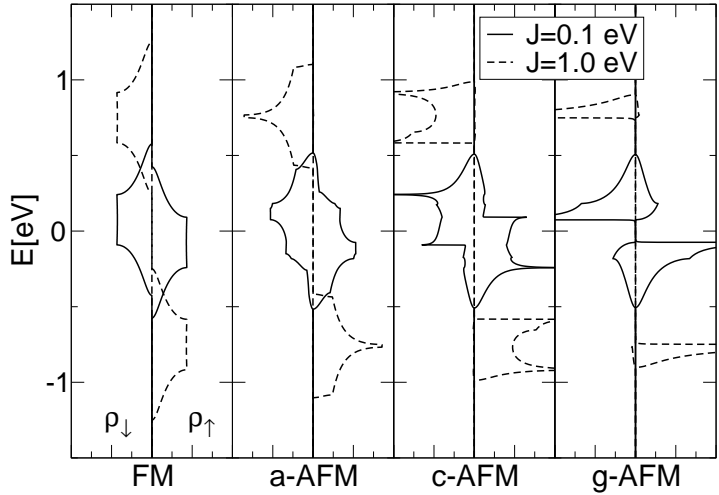}
	\caption{\label{fig:dos_MF}Sub-lattice quasi particle density of states (QDOS) of up and down
	electrons obtained from the MF-GF (\ref{eq:MF_ocpa}) for two values of local coupling $J$ 
	shown for different magnetic configurations. 
	Parameters: $S=\frac{3}{2}$ and free electron bandwidth: $W=1.0$ eV.}
    \end{center}
\end{figure}

\begin{figure}
    \begin{center}
	\includegraphics[width=.5\linewidth]{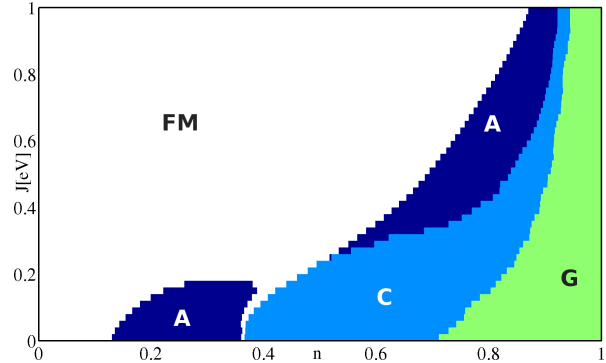}
	\caption{\label{fig:phase_MF_od} Phase diagram obtained from the MF-GF (\ref{eq:MF_ocpa})  without
    dilution regarding coupling $J$ and electron density $n$. Parameters: $S=\frac{3}{2}$ and free electron bandwidth: $W=1.0$ eV.}
    \end{center}
\end{figure}

The internal energy of the FKLM at $T=0$ is given as an integral (\ref{eq:internal_E}) over the product
of (sub-lattice) quasi-particle density of states (QDOS) times energy up to Fermi-energy.
For understanding the resulting phase-diagrams it is therefore useful to have a closer look at the
QDOS first. In figure \ref{fig:dos_MF} the sub-lattice MF-QDOS is shown for the different magnetic phases investigated.
The underlying full lattice is of simple cubic type with nearest neighbor hopping $t$ chosen such that
the bandwidth $W$ is equal to $W=1$ eV in the case of free electrons ($J=0$ eV). 
The local magnetic moment is equal to $S=\frac{3}{2}$.
We have plotted the up and down-electron spectrum separately for two different values of 
$J=0.1/1.0$ eV.
The exchange splitting $\Delta_{ex}=JS$ eV of up and down-band is clearly visible.
The decisive difference between the phases for nonzero values of $J$ is bandwidth reduction
from ferromagnetic over a, c to g-AFM phase. The reason for this behavior becomes clear by
looking at the magnetic lattices shown in figure \ref{fig:latticetypes}. In the 
ferromagnetic case an (up-)electron can move freely in all 3 directions of space without
paying any additional potential energy. In A-type anti-ferromagnetic phase the electron 
can still move freely within a plane but when moving in the direction perpendicular to the plane
it needs to overcome an energy-barrier $\Delta_{ex}$. Hence the QDOS for large values of $J$
resembles the form of 2D tight-binding dispersion. The bandwidth is reduced due to the 
confinement of the electrons. In the c-AFM phase the electron can only move freely along one 
direction and the QDOS becomes effectively one dimensional. Finally in the G-type phase the
electron in the large $J$ limit is quasi localized and the bandwidth gets very small.
This bandwidth-effect is the main reason for the phase diagram obtained as shown in figure \ref{fig:phase_MF_od}.\\
For larger $J$ ($J>0$) a typical sequence 
appears: for low band-fillings $n$ the system is always ferromagnetic and, with increasing $n$,
it becomes A-type then C-type and finally G-type anti-ferromagnetic. This behavior is 
understood easily by looking at the formula for the internal energy (\ref{eq:internal_E}) and the
MF-QDOS in figure \ref{fig:dos_MF}. Because of the bandwidth-effect discussed already, the
band-edge of the ferromagnetic state is always lowest in energy and will give therefore the
lowest internal energy for small band-occupation. 
But since the upper edges of the low energy spin-bands decrease in the order fm, a-, c-, g-AFM, the antiferromagnetic phases become eventually lower in energy for increasing band-filling. Note that for $n=1$ the lower spin sub-band is completely filled.
Therefore the bandwidth-effect is the main reason for the observed order of phases with increasing $n$. A very interesting feature can be found
in the region: $J=0.2\dots0.3$. In this region the ferromagnetic phase is directly followed by
the c-AFM phase for increasing $n$ although the a-AFM phase has a larger bandwidth than the c-AFM phase.
This can be explained by the two-peak structure of the c-AFM-QDOS. Due to the first peak at low energies 
these energies are much more weighted than in the a-AFM case and the c-AFM phase will become lower
in energy than the a-AFM phase. Since the reduction of bandwidth of the anti-ferromagnetic phases
compared to the ferromagnetic phase is more pronounced for larger values of $J$ the ferromagnetic
region is growing in this direction.\\ The paramagnetic phase does not appear for any finite $J$. Due to the down-shift of the up-spectrum of the ferromagnetic sub-lattice its internal energy will always be lower. This will change when we include correlation effects
within our MCDA treatment of the FKLM. 

\subsection{MCDA results without dilution}

\begin{figure}[htb]
 \begin{center}
   \includegraphics[width=.4\textwidth]{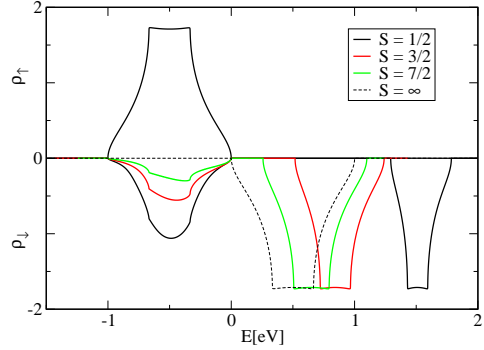}
\caption{\label{fig:qdosMCDA}QDOS of the MCDA-GF in the concentrated system for various spins $S$ ($JS=1$eV) in the ferromagnetic phase. The smaller the spin the larger is the amount of spin scattering states in the spin-down QDOS at $E\approx-\frac{J}{2}S$. The upper sub-band at $E\approx\frac{J}{2}(S+1)$ gets to lower energy for increasing spin and constant $JS$. For $S\rightarrow\infty$ the scattering states vanish and the mean-field QDOS is reproduced. Parameters: $W=1$eV , $n=0.05$}
 \end{center}
\end{figure}

\begin{figure}
    \begin{center}
	\includegraphics[width=.5\linewidth]{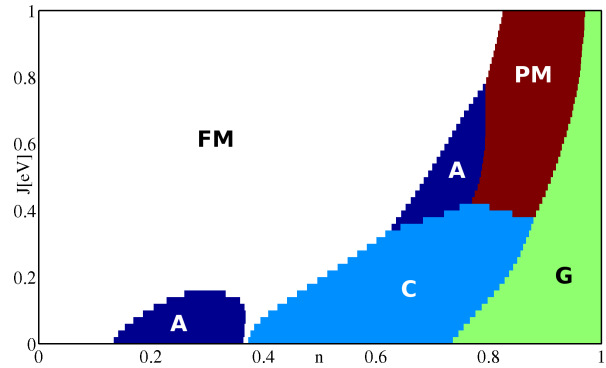}
	\caption{\label{fig:phase_MCDA_od} Phase diagram obtained from the MCDA-GF (\ref{eq:MF_ocpa})  without
    dilution. Parameters: $S=\frac{3}{2}$ and free electron bandwidth: $W=1.0$ eV.}
    \end{center}
\end{figure}

All effects that have been discussed for the mean-field results are also present in the MCDA. It means that the differences of the bandwidths and the peak structures of the single phases remain the dominant origin of differences in the internal energy. But mainly due to the spinflip GF, which vanishes for the mean-field approximation, the QDOS changes. Thus the phase diagram should be changed, too. Figure \ref{fig:qdosMCDA} shows the QDOS in the ferromagnetic phase of a concentrated lattice.
The main change compared to the MF-QDOS is the appearance of spin-down states at the energies of the spin-up band. These states can be connected to magnon emission processes. It turns out that the spectral weight of this part of the spin-down QDOS is heavily affected by the spin quantum number of the local moments. The smaller the spin the stronger are the influences of the scattering states. For very large spins the mean-field picture is reproduced (classical limit). But as it can be seen in figure \ref{fig:phase_MCDA_od} the phase diagram changes for spin $S=\frac{3}{2}$, especially at stronger couplings $J$.
Actually we get a big region of the paramagnetic phase, which was never present in the mean-field diagrams. This is plausible because in the paramagnetic regime the spin-flip processes should be important, which are not neglected by the MCDA. Thus the energy of this phase is much lower than for the mean-field approximation and can be of the same order as in the (A)FM phases. This broadening of the PM phase with increasing quantum character, was found in Ref. \cite{jochen}, too.\\
At low couplings the MF and MCDA phase diagrams are similar again, also for $S=\frac{3}{2}$. This can be easily understood if we arrange the MCDA self-energy (\ref{eq:selfMCDA}) in powers of $J$ yielding:
\begin{equation}
 M_{\sigma}(E) = -\frac{J}{2}z_{\sigma}\mean{S^z_i} +\frac{J^2}{4}M'_{\sigma}(E)\ .\label{eq:MCDAseries}
\end{equation}
We see that the self-energy splits into a MF and a "many-body" part $M'_{\sigma}(E)$. Since the MF term varies linearly with $J$ it becomes dominant at small couplings.

\subsection{Results with dilution}

\begin{figure}
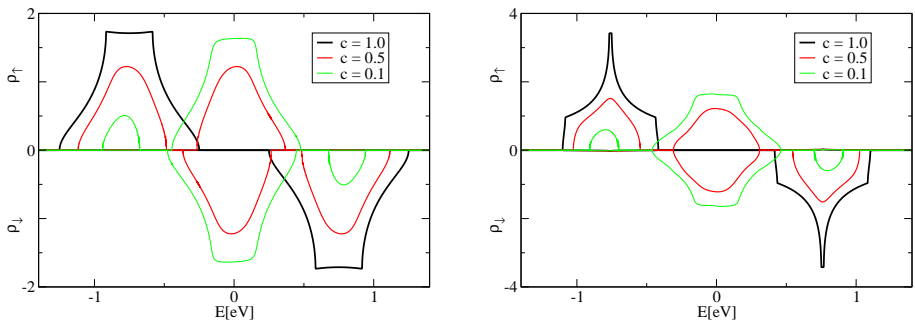

    \begin{center}
\begin{minipage}{.4\linewidth}
 	\includegraphics[width=.9\linewidth]{./figure7a.eps}
\end{minipage}
\begin{minipage}{.4\linewidth}
 	\includegraphics[width=.9\linewidth]{./figure7b.eps}
\end{minipage}
	\caption{\label{fig:dos_MF_varconz}
     MF-QDOS of the ferromagnetic (left) and A-type 
    anti-ferromagnetic (right) configuration for various concentrations $c$ of local moments.
	Parameters: $J=1.0$ eV, $S=\frac{3}{2}$ and free electron bandwidth: $W=1.0$ eV.}
    \end{center}
\end{figure}
\begin{figure}
    \begin{center}
\begin{minipage}{.05\linewidth}
  \rotatebox{90}{MF ($S=\frac{3}{2}$)}
\end{minipage}
\begin{minipage}{.44\linewidth}
\center{c=0.4}
  	\includegraphics[width=.99\linewidth]{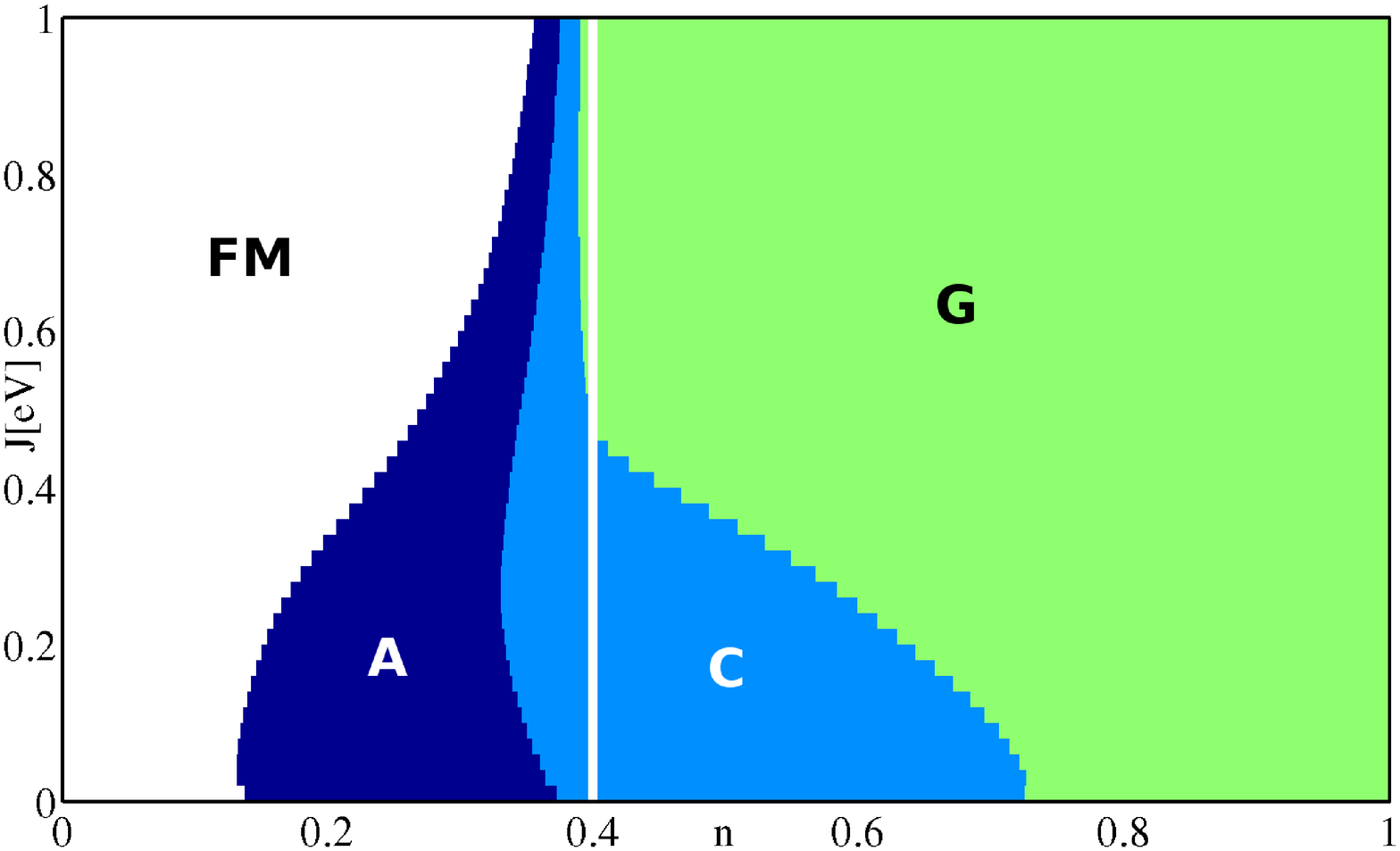}
\end{minipage}
\begin{minipage}{.44\linewidth}
\center{c=0.8}
  	\includegraphics[width=.99\linewidth]{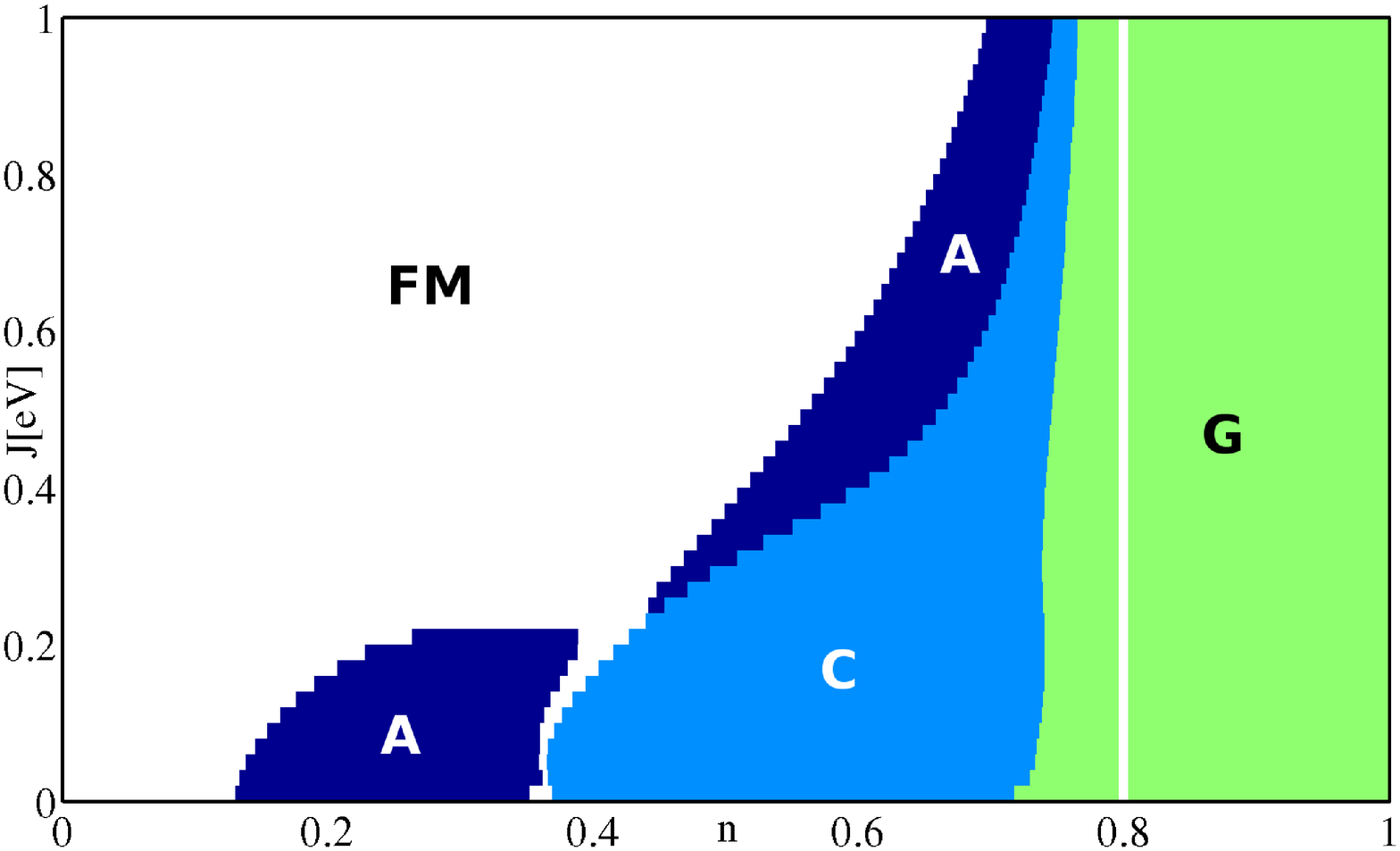}
\end{minipage}
\begin{minipage}{.05\linewidth}
 \rotatebox{90}{MCDA ($S=\frac{7}{2}$)}
\end{minipage}
\begin{minipage}{.44\linewidth}
 	  	\includegraphics[width=.99\linewidth]{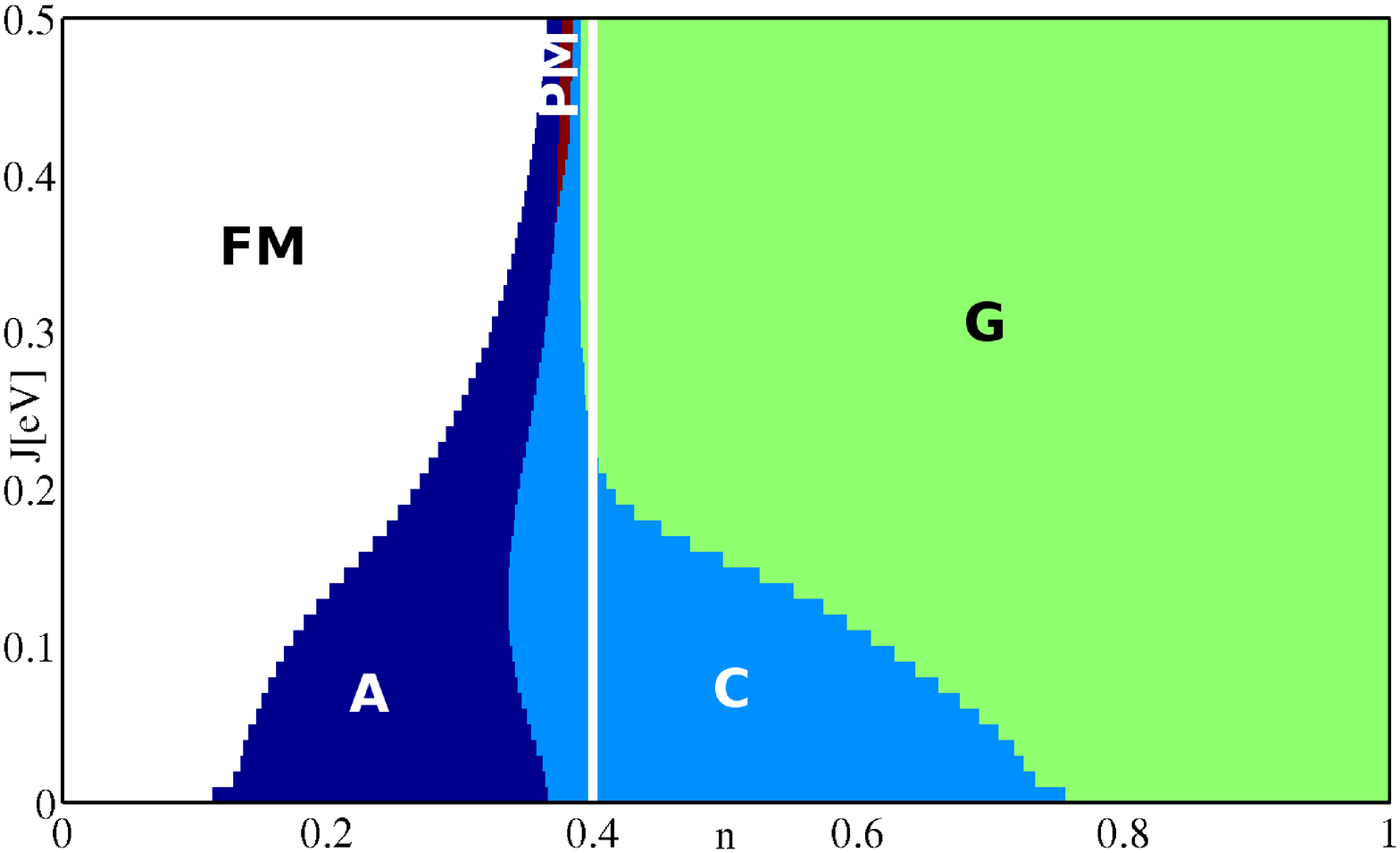}
\end{minipage}
\begin{minipage}{.44\linewidth}
 	  	\includegraphics[width=.99\linewidth]{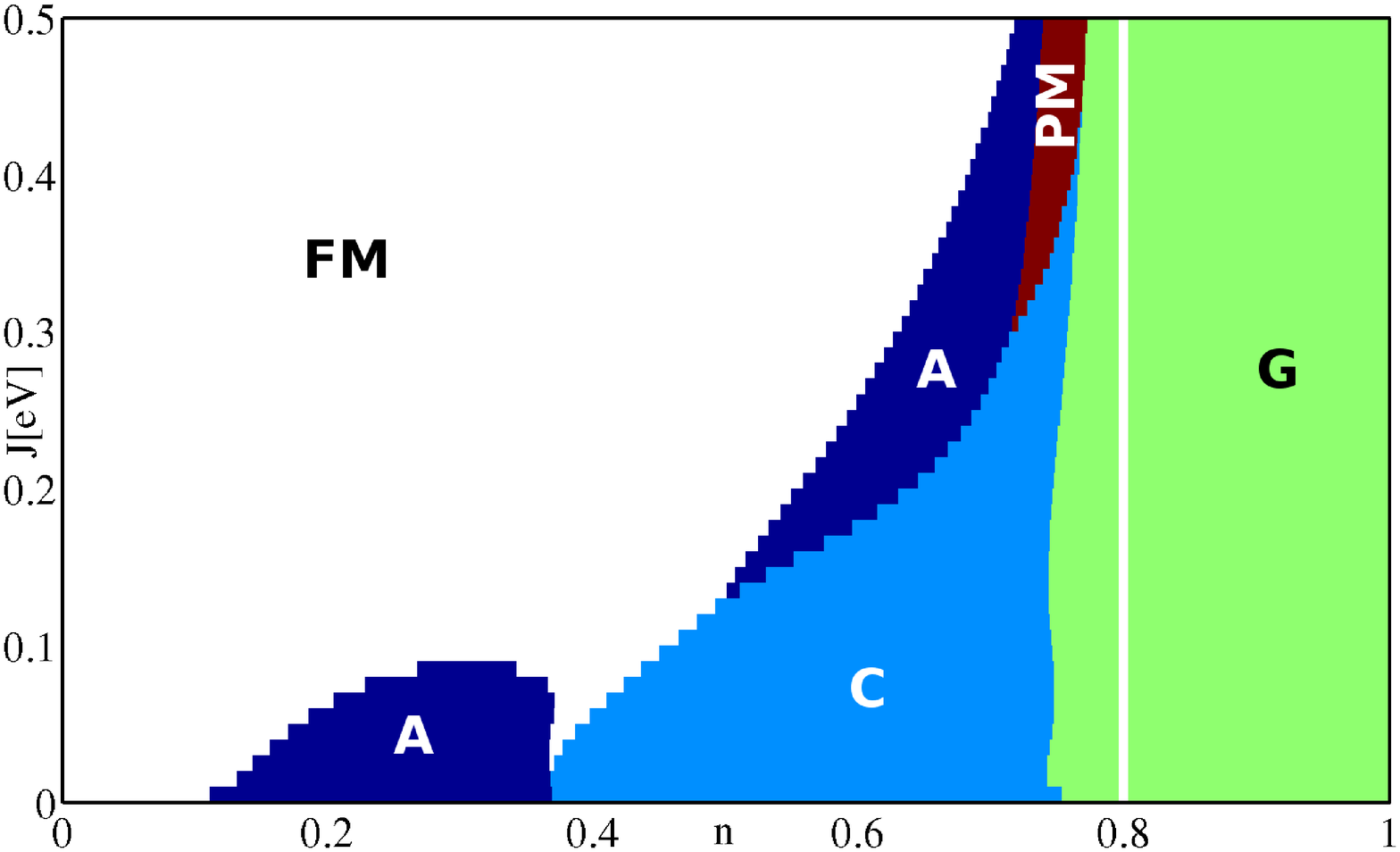}
\end{minipage}
\begin{minipage}{.05\linewidth}
 \rotatebox{90}{MCDA ($S=\frac{3}{2}$)}
\end{minipage}
\begin{minipage}{.44\linewidth}
 	  	\includegraphics[width=.99\linewidth]{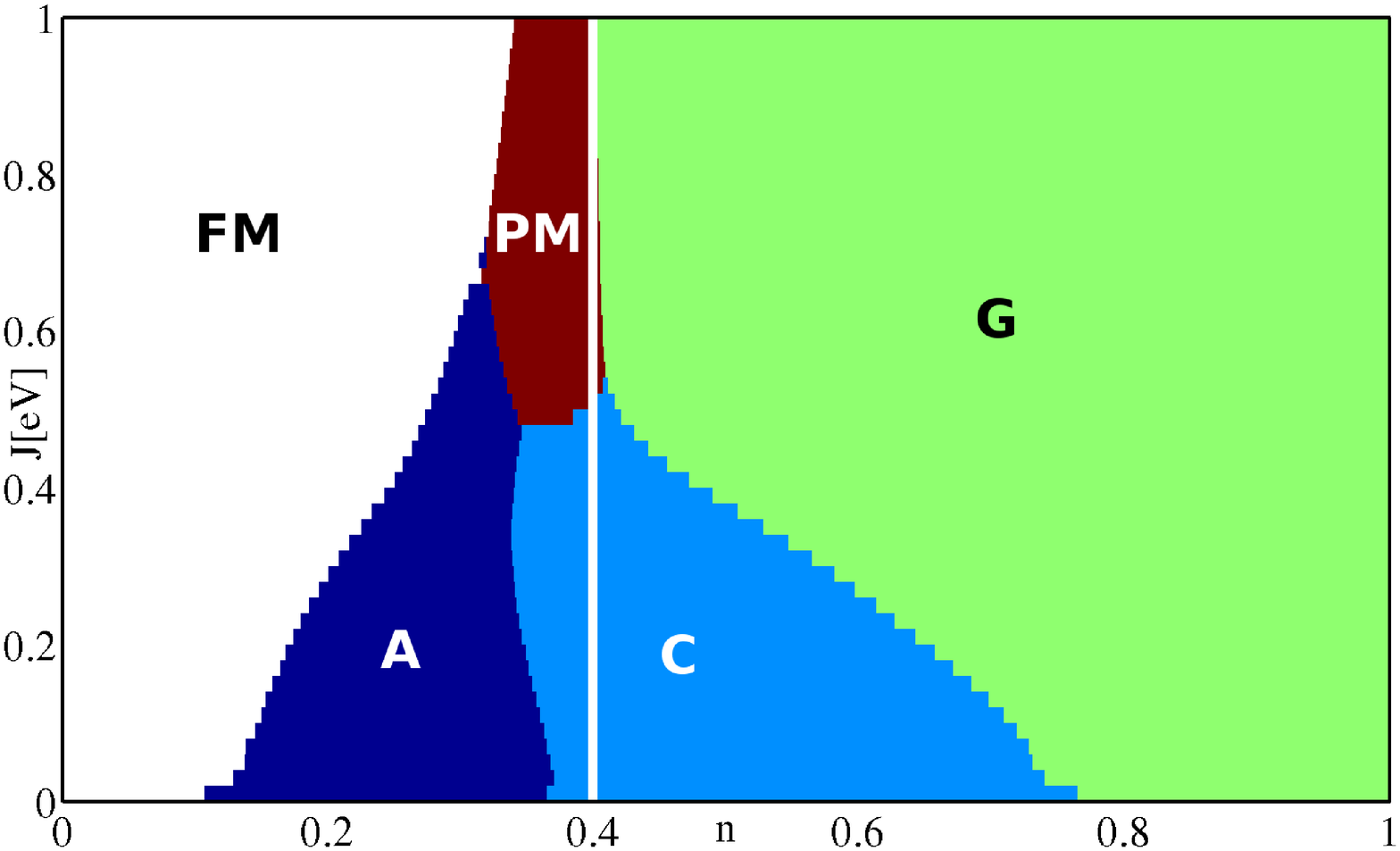}
\end{minipage}
\begin{minipage}{.44\linewidth}
 	  	\includegraphics[width=.99\linewidth]{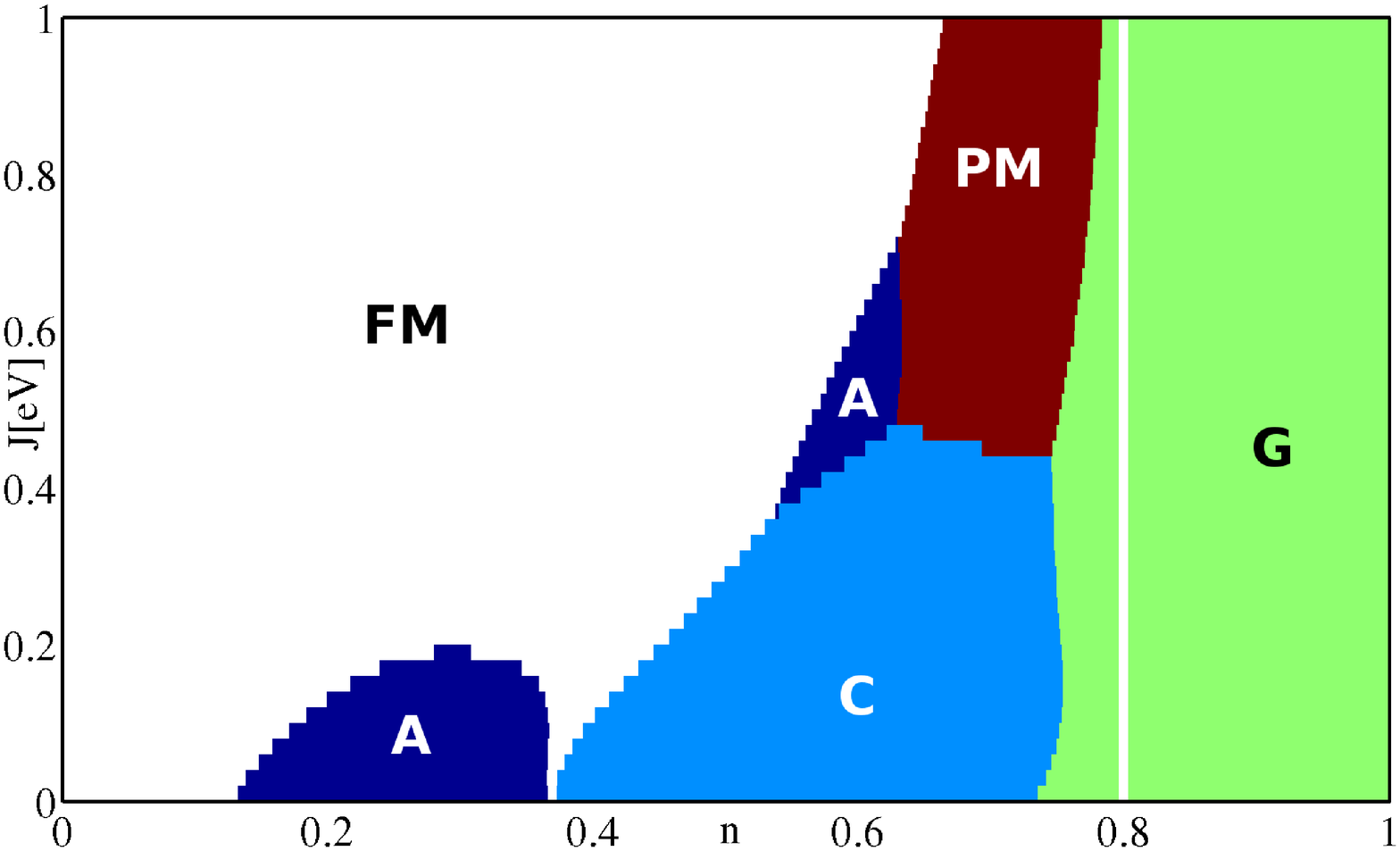}
\end{minipage}
	\caption{\label{fig:phase_MF_varconz}
     Phase diagrams with dilution for two different concentrations. \emph{left column:} $c=0.4$ \emph{right column:} $c=0.8$. The white lines show the band filling $n=c$. The two phase diagrams in the top line are calculated with the mean-field approximation. The others with the MCDA for spin $S=\frac{7}{2}$ and $S=\frac{3}{2}$, respectively. Differences between MF and MCDA results appear at large couplings $J$ and become smaller for large spins. Parameters: Free electron bandwidth: $W=1.0$ eV.}
    \end{center}
\end{figure}

After we made clear the differences and similarities of the MF and MCDA in an undiluted lattice we want to discuss the effects of dilution. These are in principle the same for both methods.  In figure \ref{fig:dos_MF_varconz} we show the MF-QDOS for the fm and a-AFM configuration at various concentrations
 $c$ of the local moment system. With increasing dilution the spectral weight of the magnetic sub-bands decreases
(proportional to $c$) and an uncorrelated band appears around the band-center of gravity $T_0=0$ eV. 
The shape of the magnetic sub-bands will still resemble the form of the undiluted case (for not too small $c$) whereas the appearing non-correlated band
is more or less the same for all magnetic configurations and only depends on the dispersion of the underlying
chemical lattice. For large exchange splitting $J$ the magnetic sub-bands are well separated
from the non-magnetic band and we can define their effective filling $n^{\mathrm{eff}}=\frac{n}{c}$. If the magnetic sub-system would be independent of the total system the phase diagrams for all concentrations $c$ would be the same according to $n^{\mathrm{eff}}$. But due to hybridization of the magnetic sub-system with the total system the antiferromagnetic phases are remarkably suppressed for $n^{\mathrm{eff}}<1$ (Figure \ref{fig:phase_MF_varconz}). This is best recognizable for the G-type AFM. For low concentrations (here $c=0.4$) this phase is almost vanishing for $\frac{n}{c}<1$ although it was present in the undiluted system for $n\gtrsim 0.9$. \\
\begin{figure}[htb]
 \begin{center}
   \includegraphics[width=.4\textwidth]{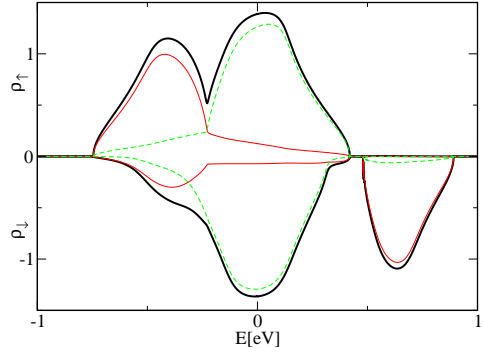}
\caption{\label{fig:qdosMCDA2}  QDOS of the MCDA-GF $G^{\mathrm{eff},\alpha\alpha}_{ii\sigma}(E)$ (bold black line), the (non-)magnetic sub-system's GF $G^{M,\alpha\alpha}_{ii\sigma}(E)$ (red line), $G^{NM,\alpha\alpha}_{ii\sigma}(E)$ (green dashed line) in a diluted system (ferromagnetic phase). For better comparing the sub-system's GF are multiplied by $c$ or $1-c$, respectively. Parameters: $c=0.4$, $J=0.5$eV, $S=\frac{3}{2}$, $W=1$eV , $n=0.05$}
 \end{center}
\end{figure}
For fillings larger than $n=c$ the non-magnetic band starts to be filled and
the phase will not change any more. Thus for $n>c$ only the g-AFM phase exists due to a completely filled lower correlated band corresponding to an effective half-filling at the correlated spectrum. From figures \ref{fig:phase_MF_od} and \ref{fig:phase_MCDA_od} we know g-AFM is just the existent phase for an undiluted lattice at $n=1$.\\ 
As seen in the case of an undiluted lattice the differences of the MF approximation and the MCDA are largest for strong couplings. That is also true in the diluted systems. But again, as for $c=1$, this is more prominent for low spins. Actually there are almost no differences in the phase diagram of the MCDA with $S=\frac{7}{2}$ compared to the MF picture but for $S=\frac{3}{2}$ a large paramagnetic phase appears (cf. figure \ref{fig:phase_MF_varconz} again).\\
The picture will change for small $J$ and small concentrations $c$. In this parameter regime the non-magnetic band
becomes broad and the exchange splitting is so small that the magnetic sub-bands are lying partly or fully
within the non-magnetic band. This can be easily seen in figure \ref{fig:qdosMCDA2} with the help of the projected GFs in (\ref{fig:qdosMCDA}).
In this case the effective filling of the magnetic sub-bands is not $n^{\mathrm{eff}}=n/c$ as in
the large $J$ limit but less since also states of the non-magnetic band are filled. Therefore the sequence of phases
with increasing $n$ becomes more and more like the undiluted case at lower $J$. This is consistent because in the limit of $J\rightarrow 0$ there is no difference between a diluted and a concentrated lattice, because both sub-systems have a vanishing potential ($\eta^{\mathrm{M}}(E)\rightarrow 0=\eta^{\mathrm{NM}}$) and can not be distinguished from each other.  For very low couplings the same sequence of phases reproduces (FM: $0\lesssim n\lesssim 0.1$, A: $0.1\lesssim n\lesssim 0.35$,C: $0.35\lesssim n\lesssim 0.75$, G: $0.75\lesssim n\leq 1$) for all concentrations. Since the MF and MCDA results are the same for low coupling due to (\ref{eq:MCDAseries}), this holds for both methods, too.\\
It is worth to make some comments about the connection of a system's $T=0$ phase diagram and its finite temperature behavior. At large couplings $J$ the energy differences between the magnetic phases are usually large (except at the borders between two phases). This implies that the respective phase is stable at moderate finite temperatures, too. On the other hand energy differences decrease when the coupling gets smaller. Thus even when an ordered phases appears in the phase diagram its critical temperature ($T_C$, $T_N$) gets very small, so that the paramagnetic phase arises at very low temperatures. In the extreme limit $J=0$ all phases have the same energy and are indistinguishable. Same is true for the limit of extreme dilution $c\rightarrow 0$. Even though the phase diagram would consist almost only of the ferromagnetic phase for $n < c$, what maybe contradicts the intuition, the Curie temperature would be very low for all couplings $J$, possibly even $T_C=0^+$.

\section{Summary and Outlook} 

We have compared the internal energy of several ordered magnetic phases and the paramagnetic phase in a diluted ferromagnetic Kondo-lattice model. This allows us to make conclusions about the existence of the various magnetic phases at $T=0$ as a function of the coupling $J$, the band filling $n$ and the concentration of the magnetic atoms $c$. To get the internal energy we applied two approximations
to treat the ferromagnetic Kondo-lattice model. The expressions for the Green's function's self-energy of the (concentrated) FKLM defined effective potentials and therewith a dynamical alloy analogy which we included into a coherent potential approximation. We used this to handle the problem of the diluted disordered system. As a first approximation for the FKLM we used a relatively simple mean-field decoupling. But already this simple ansatz gave insight in the main mechanisms of stabilizing a definite phase at a given parameter set. A more accurate moment conserving decoupling approach leads to the statement that the mean-field results are appropriate at low couplings or large spin quantum numbers $S$. For low spins the system has a more pronounced quantum character and we have distinct deviations from the mean-field picture. Both methods made clear that for very low couplings $J$ and arbitrary concentrations $c$ all magnetic phases appear at the same band fillings as in an undiluted system.\\
It is surprising that the simple mean-field ansatz leads, under certain conditions, to relatively good results at $T=0$. This should change when we go to finite temperatures. The paramagnetic phase is treated completely incorrectly in the MF approximation, due to the importance of fluctuations at $\mean{S^z}\neq S$. Thus it is not sufficient to use it  for, e.g., calculating Curie temperatures. To do this one has to use more accurate many-body methods. Expanding this theory for finite temperatures is very important to make predictions about possible candidates for dilute magnetic semiconductors at room temperature. To do this, the introduced pure KLM is surely not enough to describe DMS properly. For example, the magnetic and non-magnetic sites consist of different atoms and therefore atomic levels. This results in additional changes of the according density of states and will be done in forthcoming work.

\begin{appendix}
\section{\label{app:coeff}Coefficients and expectation values of the MCDA}

The moments of the various Green's functions $X^A_{ij\sigma}=\green{A_{i\sigma}}{c^+_{j\sigma}}$, where $A_{i\sigma}$ is an arbitrary operator combination, are defined by
\begin{equation*}
 M^{(n)}_{ij\sigma}( X^A_{ij\sigma})= \mean{\Big[ \underbrace{[\dots[A_{i\sigma},H]_-\dots,H]_-}_{(n-p)-\mathrm{fold}}, \underbrace{[H,\dots[H,c^+_{j\sigma}]_-\dots]_-}_{(p)-\mathrm{fold}}\Big]_+}\ .\label{eqmomente}
\end{equation*}
Explicit calculation of the moments yields the coefficients of the MCDA as
\begin{eqnarray*}
\fl \alpha^{(1)}_{\sigma} &= 0\\
\fl \beta^{(1)}_{\sigma} &=\frac{ 3z_{\sigma}\mean{S^{\bar\sigma}_iS^{\sigma}_i}+2\mean{S^z_i} + \mean{S^{\sigma}_iS^{\bar\sigma}_iS^z_i} -2z_{\sigma} S(S+1)(1-\mean{n_{i\bar\sigma}}) +4\Delta_{\bar\sigma} - 3z_{\sigma}\mu_{\bar\sigma} - \eta_{\sigma}}{\mean{S^{\bar\sigma}_iS^{\sigma}_i}-\gamma_{\sigma}+2z_{\sigma}\Delta_{\bar\sigma}}\\
\fl \alpha^{(2)}_{\sigma} &= 0\\
\fl \beta^{(2)}_{\sigma} &=\frac{\mean{S^{\bar\sigma}_iS^{\sigma}_iS^z_i}-\mean{S^{\bar\sigma}_iS^{\sigma}_i}\mean{S^z_i}+2\eta_{\sigma}}{\mean{(S^z_i)^2}-\mean{S^z_i}^2-\gamma_{\sigma}} \\[.25cm]
 \fl \alpha^{(3)}_{\sigma} &= -\gamma_{\sigma}\\
\fl \beta^{(3)}_{\sigma} &=\mean{n_{i\sigma}}\\
\fl  \alpha^{(4)}_{\sigma} &= \Delta_{\bar\sigma}\\
\fl \beta^{(4)}_{\sigma} &=\mean{n_{i-\sigma}}
\end{eqnarray*}
where we suppressed the index $\alpha$ at every site index $i$ for readability. We have abbreviated
\begin{eqnarray*}
 \Delta_{\sigma}&=& \mean{S^z_i n_{i\sigma}}\\
\gamma_{\sigma}&=&\mean{S^{-\sigma}_i c^+_{i\sigma}c_{i-\sigma}}\\
\eta_{\sigma}&=& \alpha^{(1)}_{\sigma}\mean{n_{i\sigma}}+\beta^{(1)}_{\sigma}\gamma_{\sigma}\\
\mu_{\sigma}&=&(\alpha^{(2)}_{\sigma}+\mean{S^{-\sigma}_iS^{\sigma}_i})\mean{ n_{i\sigma}} +\beta^{(2)}_{\sigma}(\Delta_{\sigma}-\mean{S^z_i n_{i\sigma}})\\
 \nu_{\sigma} &=& \alpha^{(3)}_{\sigma}\mean{n_{i\sigma}}+\beta^{(3)}_{\sigma}\gamma_{\sigma}\\
\vartheta_{\sigma}&=& \alpha^{(4)}_{\sigma}\mean{n_{i\sigma}}+\beta^{(4)}_{\sigma}\Big(\Delta_{\sigma}-\mean{S^z_i}\mean{n_{i\sigma}}\Big)\ .
\end{eqnarray*}
The prefactors appearing in the equations of motion are
\begin{eqnarray*}
A_{\Gamma\sigma} &= z_{\sigma}\mean{(\delta S_i^z)^2}+z_{\sigma}\alpha_{3\sigma}\\
B_{\Gamma\sigma} &= z_{\sigma}\mean{S_i^z} -1\\
C_{\Gamma\sigma} &= -z_{\sigma} - \mean{S_i^z}+\beta_{1\sigma}+z_{\sigma}\beta_{3\sigma}\\
A_{F\sigma} &=\alpha_{3\sigma}+2z_{\sigma}\alpha_{4\sigma}\\
B_{F\sigma} &=2z_{\sigma}\beta_{4\sigma}\\
C_{F\sigma} &= -z_{\sigma}\beta_{1\sigma}+\beta_{3\sigma}\ .
\end{eqnarray*}
The spin's expectation values are trivial for ferromagnetic saturation and for paramagnetism we get them from the solution by Callen \cite{callen}. This yields
\begin{eqnarray*}
 \mathrm{saturation:}& &\\
 \mean{(S^z_i)^n}&=&S^n\\
\mean{S^{-\sigma}_i S^{\sigma}_i}&=&S-z_{\sigma}S\\
\mean{S^{-\sigma}_i S^{\sigma}_iS^z_i}&=&(S-z_{\sigma}S)S\\
 \mathrm{paramagnetism:}& &\\
 \mean{S^z_i}&=&\mean{(S^z_i)^3}=0\\
 \mean{(S^z_i)^2}			&=& \frac{1}{3}S(S+1)\\
 \mean{S^{-\sigma}_i S^{\sigma}_i}&=&\frac{2}{3}S(S+1)\\
\mean{S^{-\sigma}_i S^{\sigma}_iS^z_i}&=&-z_{\sigma}\mean{(S^z_i)^2}\ .
\end{eqnarray*}
The expectation values of the magnetic subsystem can be calculated via the spectral theorem with the according Fourier-transformed Green's functions
\begin{eqnarray*}
 \mean{n_{i\alpha\sigma}}&= -\frac{1}{N\pi}\sum_{\mathbf k}\int dE f_-(E,\mu)\mathrm{Im} G^{\mathrm{M}}_{\mathbf k\alpha\sigma}(E)\\
 \mean{S^z_i n_{i\alpha\sigma}}&= \mean{S^z_i}\mean{n_{i\sigma}}-\frac{1}{N^2\pi}\sum_{\mathbf k\mathbf q}\int dE f_-(E,\mu)\mathrm{Im} \Gamma^{\mathrm{M}}_{\mathbf k\mathbf q\alpha\sigma}(E)\\
\mean{S^{-\sigma}_i c^+_{i\alpha\sigma}c_{i\alpha-\sigma}}&=- \frac{1}{N^2\pi}\sum_{\mathbf k\mathbf q}\int dE f_-(E,\mu)\mathrm{Im} F^{\mathrm{M}}_{\mathbf k\mathbf q\alpha\sigma}(E) \ .
\end{eqnarray*}
Note that the chemical potential $\mu$ within the Fermi functions $f_-(E,\mu)$ is derived from the whole diluted system via the condition
\begin{equation*}
 n= -\frac{1}{N\pi}\sum_{\mathbf k\sigma}\int dE f_-(E,\mu)\mathrm{Im} G^{\mathrm{eff}\alpha\alpha}_{\mathbf k\sigma}(E)\ .
\end{equation*}
This means that only the electron number of the total system is conserved, while the number of electrons in the magnetic sub-system can change.

\end{appendix}

\section*{References}
\bibliography{citations}
\bibliographystyle{unsrt}
\end{document}